\documentclass{emulateapj}

\usepackage{color}

\slugcomment{submitted to ApJ}

\shorttitle{Optical light from NGC 5408 X-1}
\shortauthors{Kaaret \& Corbel}

\begin{document}

\title{A Photoionized Nebula Surrounding and Variable Optical Continuum
Emission from the Ultraluminous X-Ray Source in NGC 5408}

\author{Philip Kaaret\altaffilmark{1,2} and St\'ephane
Corbel\altaffilmark{1}}

\altaffiltext{1}{AIM$-$Unit\'e Mixte de Recherche CEA$-$CNRS and
Universit\'e Paris VII$-$UMR 7158, CEA Saclay, Service d'Astrophysique,
F$-$491191 Gif sur Yvette, France}

\altaffiltext{2}{Department of Physics and Astronomy, University of Iowa, 
Van Allen Hall, Iowa City, IA 52242, USA}

\begin{abstract}

We obtained optical spectra of the counterpart of the ultraluminous
X-ray source NGC 5408 X-1 using the FORS spectrograph on the VLT. The
spectra show strong high excitation emission lines, He{\sc ii}
$\lambda$4686 and [Ne{\sc V}] $\lambda$3426, indicative of X-ray
photoionization.  Using the measured X-ray spectrum as input to a
photoionization model, we calculated the relation between the He{\sc ii}
and X-ray luminosities and found that the He{\sc ii} flux implies a
lower bound on the X-ray luminosity of $3 \times 10^{39} \rm \, erg \,
s^{-1}$.  The [Ne{\sc v}] flux requires a similar X-ray luminosity. 
After subtraction of the nebular emission, the continuum appears to have
a power-law form with a spectral slope of $-2.0^{+0.1}_{-0.2}$.  This is
similar to low-mass X-ray binaries where the optical spectra are
dominated by reprocessing of X-rays in the outer accretion disk.  In one
observation, the continuum, He{\sc ii} $\lambda$4686, and [Ne{\sc V}]
$\lambda$3426 fluxes are about 30\% lower than in the other five
observations.  This implies that part of the line emission originates
within one light-day of the compact object.  Fitting the optical
continuum emission and archival X-ray data to an irradiated disk model,
we find that  $(6.5 \pm 0.7) \times 10^{-3}$ of the total bolometric
luminosity is thermalized in the outer accretion disk.  This is
consistent with values found for stellar-mass X-ray binaries and larger
than expected in models of super-Eddington accretion flows.  We find no
evidence for absorption lines that would permit measurement of the
radial velocity of the companion star.

\end{abstract}

\keywords{black hole physics -- galaxies: individual: NGC 5408
galaxies: stellar content -- X-rays: galaxies -- X-rays: black holes}

\section{Introduction}

``Ultraluminous X-ray sources'' (ULXs) are variable X-ray sources in
external galaxies with luminosities (if the emission is isotropic)
greatly exceeding those of Galactic X-ray binaries.  The irregular
variability, observed on time scales from tenths of seconds to years,
suggests that ULXs are binary systems containing a compact object.  The
compact objects in these binaries are either stellar-mass black holes in
an unusual state with very high accretion rates or intermediate-mass
black holes.  It is of significant interest to determine the nature of
the compact objects in ULXs.

Beyond the luminosities, there are two lines of X-ray evidence
suggesting that ULXs harbor intermediate mass black holes: X-ray
spectral evidence for accretion disks cooler than those in stellar-mass
black hole X-ray binaries \citep{Kaaret03} and quasiperiodic
oscillations (QPOs) at lower frequencies than found in stellar-mass
black hole X-ray binaries at similar luminosities \citep{Strohmayer03}.
\citet{Feng07a} recently found that the putative cool disk in one ULX,
NGC 1313 X-2, does not follow the relation that the luminosity is
proportional to temperature to the fourth power, $L \propto T^4$, as
expected for accretion disk emission, ruling out this interpretation for
that source.  This finding casts doubt on the X-ray spectral evidence at
least for that source.  In contrast, the X-ray timing evidence appears
robust with the QPO frequencies following the relations found in X-ray
binaries \citep{Feng07b}.  QPOs have been definitively detected from
only two ULXs: one in M82 and NGC 5408 X-1
\citep{Strohmayer03,Strohmayer07}.  Further, NGC 5408 X-1 is surrounded
by a powerful radio nebula \citep{Lang07} which requires an extremely
energetic outflow from the compact object.  NGC 5408 X-1 is one of the
best intermediate mass black hole candidates amongst the ULXs.


Optical spectroscopy of ULXs provides, perhaps, the best means to better
understand the nature of the ULXs.  Indeed, the most definitive
constraint on the compact object mass would come from optical
measurement of the mass function of the companion star.  The ULX in M82
may lie in a superstar cluster and, in any event, lies in too crowded
and obscured a region to permit extraction of an optical spectrum of the
ULX binary \citep{Kaaret01,Kaaret04a}.  In contrast, NGC 5408 X-1 lies
in a relatively uncrowded region and is only weakly obscured.  We
recently identified the optical counterpart to NGC 5408 X-1
\citep{Lang07}.  Thus, NGC 5408 X-1 is the best intermediate mass black
hole candidate ULX for which optical spectroscopy of the binary system
can be performed.

Optical emission can arise from a companion star, an accretion disk
surrounding the compact object, or a surrounding nebula.  If the light
arises from the companion star, then one would expect continuum emission
and, for most spectral types, a series of absorption lines.  Measurement
of the continuum shape would help constrain the spectral type of the
companion star.  Detection of spectral lines from the companion star
would enable us to determine the star's spectral type, thereby
constraining the evolutionary history of the binary.  Repeated
observations would enable measurement of the mass function and a lower
bound on the compact object mass.

If the optical emission arises from the accretion disk, then we would
expect to see a blue or hot continuum and emission lines. 
High-excitation lines, particularly He{\sc ii}, would give direct
information about the nature of the accretion flow and the incident
X-ray flux.  These lines can provide information about the nature of the
binary system and an estimate of the true X-ray luminosity.  In
addition, since the accretion disk moves with the compact object, the
emission line radial velocity represents the radial velocity of the
compact object and thus provides a means to constrain the mass of the
compact object.  This technique was demonstrated by \citet{Hutchings87}
for the persistent black hole candidate LMC X-1.

If the optical light arises from a surrounding nebula then the nebula
could be either shock-powered, in which case low excitation lines would
be expected, or X-ray photoionized, in which case high excitation lines
would be expected.  Forbidden lines can produced in a photoionized
nebula but not in the high densities of an accretion disk.  Nebulae may
also produce continuum emission.

We obtained several observations of NGC 5408 X-1 using the FORS
instrument on the VLT and analyzed archival X-ray data from XMM-Newton. 
The observations and data reduction are described in \S 2. The results
are described in \S 3 and discussed in \S 4.

\begin{figure}
\centerline{\includegraphics[angle=0,width=3.25in]{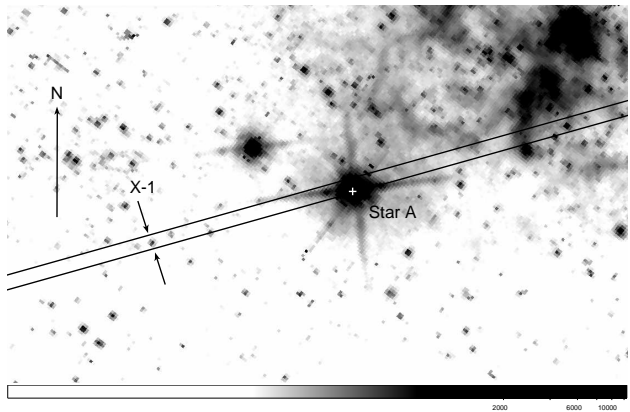}}
\caption{Image of NGC 5408 in the F606W filter (V-band) from the WFPC2
on HST.  The position of the optical counterpart of NGC 5408 X-1, the
position of star A, and the FORS1 slit ($0.51\arcsec$ wide) are marked. 
The arrow is $4\arcsec$ long and points north.}  \label{hstimage} 
\end{figure}

\begin{deluxetable}{llcc} \tabletypesize{\scriptsize}
\tablecaption{VLT/FORS Observations of NGC 5408 X-1 \label{obstable}}
\tablewidth{0pt} 
\tablehead{\colhead{Observation} & \colhead{Start} & \colhead{Duration} & \colhead{Air mass}}
\startdata 
OB1 & 2008-04-08 04:59:04 & 3$\times$850~s & 1.050 \\
OB2 & 2008-04-08 05:46:04 & 3$\times$850~s & 1.049 \\
OB3 & 2008-04-08 06:45:13 & 3$\times$850~s & 1.095 \\
OB4 & 2008-04-09 05:46:17 & 3$\times$850~s & 1.051 \\
OB5 & 2008-04-10 05:08:33 & 3$\times$850~s & 1.046 \\
OB6 & 2008-04-10 05:54:29 & 3$\times$850~s & 1.056 \\
\enddata 
\tablecomments{The table lists VLT/FORS observations of
NGC 5408 and includes the date and UT time of the start of the
observation, the number of integrations and the duration of each
integration, and the average airmass during each observation.} 
\end{deluxetable}

\newpage
\section{Observations and Analysis}

\subsection{Optical}

The Very Large Telescope (VLT) was used for the observations because of
the dimness of the optical counterpart of NGC 5408 X-1, $V=22.4$, and
the need to obtain moderately high resolution spectra.  NGC 5408 X-1
transits nearly overhead at Cerro Paranal.  We chose to use the FOcal
Reducer and Spectrograph 1 \citep[FORS1]{Appenzeller98}  which has a
blue sensitive CCD  because we are primarily interested in lines that
would be produced in a photoionized nebula or accretion disk and many of
the important lines for such systems are in the blue range.  FORS1 is
mounted on the VLT unit telescope 2 (UT2). We used the GRIS\_600B grism
which covers a wavelength range of 3350~\AA\ to 6330~\AA\ with a
dispersion of 0.75~\AA/pixel and a spectral resolution of
$\lambda/\Delta\lambda = 780$ at the central wavelength.  We used a
$0.51\arcsec$ slit and 2$\times$2 binning of the pixels with high gain
readout and standard resolution.

The spectrograph slit was oriented so that the closest neighbors of the 
ULX counterpart are best separated along the slit.  The slit also
covered a bright star, `star A', that appears in the 2MASS catalog
\citep{2MASS} at a position of $\rm R.A. = 14^h03^m18\fs97$, $\rm Decl.
= -41\arcdeg22\arcmin56\farcs6$ (J2000).  Since the ULX optical
counterpart is quite dim, star A was used for acquisition.  We checked
the slit position by correcting the astrometry of the FORS acquisition
image using the stars in the 2MASS catalog.  Fig.~\ref{hstimage} shows
an HST image of the field \citep{Kaaret04b} with the reconstructed slit
position superimposed.  The optical counterpart of NGC 5408 X-1
identified in \citet{Lang07} appears to be well placed in the slit.

Multiple observations of NGC 5408 X-1 were obtained over three
successive nights, see Table~\ref{obstable}.  The data reduction was
performed using the Image Reduction and Analysis Facility (IRAF)
\citep{Tody93}.  Starting with the raw data, we created bias and lamp
flat-field images using exposures obtained on the 3 nights of
observations.  These corrections were applied to the spectrum images.
Each observation block (OB) consisted of three integrations of 850~s
each with an offset of 12 pixels along the spatial axis between
successive integrations.  We analyzed the data by combining the 3
two-dimensional spectra in each observation block in order to remove bad
pixels and cosmic ray hits.  We used the {\tt xregister} task in IRAF to
caculate the offsets needed to align the images and then the {\tt
imcombine} task to average the images, applying the {\tt ccdclip} method
to remove cosmic rays and bad pixels.  We also combined the entire set
of 18 spectra into one using the same procedure.

The dimness of the continuum emission from the optical counterpart of
the ULX made finding a trace of the spectrum unreliable.  Instead, we
first traced the spectrum of star A and then used that as a reference
for the trace of the ULX spectrum.  The trace position on the spatial
axis varied by less than 0.5 pixels along the length of the dispersion
axis.  The trace for the ULX counterpart was centered using the profile
of the He{\sc ii} $\lambda$4686 emission line.  Wavelength lamp and
standard star (EG247) spectra were used to wavelength and flux calibrate
the spectra. Extinction corrections were applied using the CTIO
extinction tables available in IRAF.

The spectra were searched for emission and absorption lines using the
{\tt splot} task in IRAF.  A number of emission lines, discussed below
were found.  No stellar absorption lines were detected.  There are some
weak absorption features that are due to subtraction of night sky
background lines at 5580~\AA\ and just above the He {\sc i} line at
5885~\AA.  The latter night sky line makes measurement of the He {\sc i}
$\lambda$5876 flux unreliable.  After correction for the redshift of NGC
5408, $z = 0.00168785$, the lines are all within 1~\AA\ of the expected
values except for [Ne{\sc v}] $\lambda$3426 which differs by 3~\AA. 
However, [Ne{\sc v}] is at a shorter wavelength than the shortest
available in the wavelength calibration lamp spectra, 3650~\AA, and we
believe that the offset is due to extrapolation of the calibration.

\subsection{X-ray}
\label{xray}

To determine the spectrum of X-rays from the ULX, we extracted the
longest publicly available observation of NGC 5408 from the XMM-Newton
archive.  The observation was performed on Jan 13, 2006 and had a
duration of 133~ks \citep{Strohmayer07}.  We performed the standard XMM
data filtering, including removal of data from times when background
flares were present using SAS version 8.0.0.  We selected only single
pixel events in order to obtain the best energy resolution, since
adequate photon statistics were available, and binned the X-ray spectrum
in the 0.3--9.55~keV band to have at least 25 counts per bin.  We fitted
the spectrum using XSPEC version 12.5.0.

Following \citet{Kaaret04b}, we chose to use a Comptonization model,
{\tt compps} \citep{Poutanen96}, including multicolor blackbody emission
from an accretion disk because we must extrapolate the X-ray spectrum to
energies below the minimum observed X-ray energy in order to provide a
useful input to the photoionization modeling described below.  We used
the {\tt tbabs} model to account for interstellar absorption along the
line of sight.  We found evidence for an emission line or the
combination of a unresolved group of emission lines near 0.9~keV and
included a Gaussian to fit that component.  Our best fitted model gave a
$\chi^2$ of 243.2 for 219 degrees of freedom with a column density $N_H
= (7.07 \pm 0.03) \times 10^{20} \rm \, cm^{-2}$, an inner disk
temperature $kT = 172 \pm 8 \rm \, eV$, and a Comptonization electron
temperature of $kT = 74 \pm 8 \rm \, keV$ and optical depth of $0.43 \pm
0.07$.  We note that the electron temperature and optical depth depend
on the specific geometry and form of the electron distribution used. 
The Gaussian emission had a centroid of $0.91 \pm 0.02 \rm \, keV$, a
width of $67 \pm 20 \rm \, eV$, and an equivalent width of 31~eV.  This
emission component may represent a thermal plasma, as previously
suggested for this source \citep{Strohmayer07} and also seen from some
other ULXs \citep{Feng05}.  However, using the {\tt apec} thermal plasma
model instead of the single Gaussian produces a worse fit, $\chi^2$ of
263.3 for 219 degrees of freedom.  Thus, interpretation of the line is
not clear and higher resolution spectra will likely be required to
understand the origin of the emission.  Correcting for absorption and
integrating the model from 1~eV to 100~keV, the total luminosity is $1.0
\times 10^{40} \rm \, erg \, s^{-1}$.

We note that our X-ray and optical observations are not simultaneous;
they are separated by two years.  However, NGC 5408 X-1 does not appear
to be highly variable in X-rays.  The X-ray light curve constructed by
\citet{Kaaret03} spanning 22 years with multiple observatories, showed a
ratio between the maximum and minimum observed fluxes of 1.4.  Those
fluxes ranged from 7\% below to 30\% above our measured flux.  Five
Chandra observations carried out over 3 years showed a variability of
12\% and no significant changes in spectral shape.

\begin{figure}
\centerline{\includegraphics[angle=0,width=3.0in]{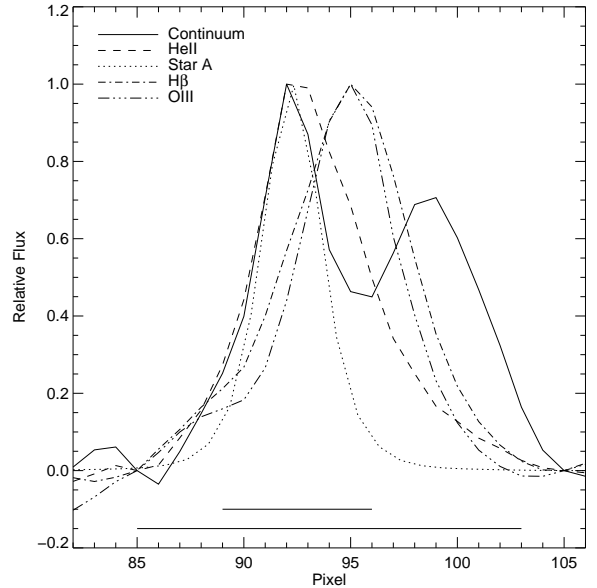}}
\caption{Spatial profiles of emission from the ULX.  The solid curve is
continuum emission in the band from 4370~\AA\ to 4680~\AA, the peak near
pixel 92 is due to the optical counterpart of the ULX and the peak near
pixel 99 is due to another star.  The dashed curve is the He{\sc ii}
emission line.  The dotted curve is the continuum emission from star A,
shifted to align with the ULX profile and shown for comparison.  The
continuum emission from the ULX appears point like, while the He{\sc ii}
is slightly extended.  The dot-dashed curve is H$\beta$.  The
dot-dot-dot-dashed curve is [O{\sc iii}].  Both of these emission lines
are broader than and displaced from the ULX continuum.  Near the bottom
of the plot, the upper solid horizontal line shows the spatial region
used for the narrow trace and the lower horizontal line shows the wide
trace.  These are the spatial regions used for extraction of spectra as
discussed in the text.  Each pixel is $0.25\arcsec$.}  \label{profile}
\end{figure}

\begin{figure*}
\centerline{\includegraphics[angle=0,width=6.0in]{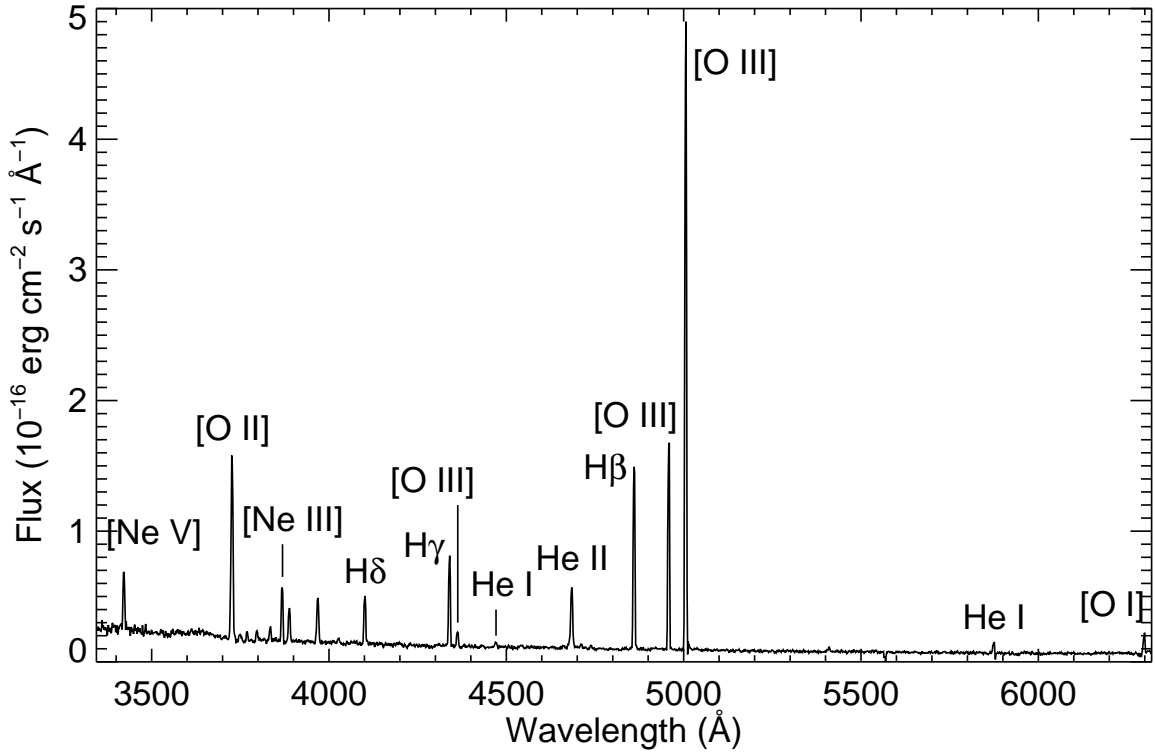}}
\caption{Spectrum of the optical counterpart of NGC 5408 X-1.  The
dereddened flux is plotted versus redshift-corrected wavelength.} 
\label{specall}  \end{figure*}

\begin{deluxetable}{lccc} \tabletypesize{\scriptsize}
\tablecaption{Optical emission lines from NGC 5408 X-1\label{lines_table}}
\tablewidth{0pt} 
\tablehead{\colhead{Identification} & \colhead{Wavelength (\AA)} &\colhead{Flux} & \colhead{FWHM (\AA)}}
\startdata 
$[$Ne~V]      $\lambda$3426 &  3428.9 &   315 $\pm$ 21 &   5.2 $\pm$ 0.2 \\
$[$O~II]      $\lambda$3727 &  3734.4 &  1103 $\pm$ 40 &   6.2 $\pm$ 0.1 \\
H$\iota$      $\lambda$3771 &  3777.0 &    46 $\pm$ 11 &   4.5 $\pm$ 1.1 \\
H$\theta$     $\lambda$3798 &  3804.8 &    56 $\pm$ 12 &   5.4 $\pm$ 1.1 \\
H$\eta$       $\lambda$3835 &  3842.4 &    71 $\pm$ 11 &   5.1 $\pm$ 0.8 \\
$[$Ne~III]    $\lambda$3869 &  3875.9 &   285 $\pm$ 15 &   4.9 $\pm$ 0.2 \\
H$\zeta$, He {\sc i} $\lambda$3889 &  3896.1 &   188 $\pm$ 14 &   5.5 $\pm$ 0.3 \\
H$\epsilon$, $[$Ne {\sc iii}] $\lambda$3970 &  3976.6  &   260 $\pm$ 16 &   5.8 $\pm$ 0.3 \\
H$\delta$     $\lambda$4102 &  4109.5 &   268 $\pm$ 14 &   5.4 $\pm$ 0.2 \\
H$\gamma$     $\lambda$4340 &  4348.6 &   471 $\pm$ 15 &   5.1 $\pm$ 0.1 \\
$[$O~III]     $\lambda$4363 &  4371.2 &    81 $\pm$ 10 &   5.3 $\pm$ 0.7 \\
He~II         $\lambda$4686 &  4693.7 &   382 $\pm$ 12 &   6.3 $\pm$ 0.2 \\
H$\beta$      $\lambda$4861 &  4869.9 &  1000 $\pm$  8 &   5.2 $\pm$ 0.1 \\
$[$O~III]     $\lambda$4959 &  4967.8 &  1083 $\pm$  5 &   5.1 $\pm$ 0.1 \\
$[$O~III]     $\lambda$5007 &  5015.7 &  3226 $\pm$  4 &   5.1 $\pm$ 0.1 \\
He~I          $\lambda$5876 &  5885.3 &    57 $\pm$  6 &   4.9 $\pm$ 0.8 \\
$[$O~I]       $\lambda$6300 &  6310.5 &   101 $\pm$  5 &   5.2 $\pm$ 0.5 \\
\enddata

\tablecomments{The table gives the dereddened fluxes of emission lines
relative to the flux of H$\beta$ = 1000.  The dereddened H$\beta$ flux
is $(8.6 \pm 0.9)\times 10^{-16} \, \rm erg \, cm^{-2} \, s^{-1}$.}
\end{deluxetable}

\section{Results}

Figure~\ref{profile} shows profiles along the spatial direction of the
H$\beta$ emission line, the He{\sc ii} emission line, and integrated
flux over the 4370-4680~\AA\ continuum band.  The profiles were
extracted from the combination of the entire set of 18 spectra.  The
spatial profile of the continuum flux from Star A in the same band,
shifted to match the profile of the ULX, is shown for comparison.  The
continuum emission from the ULX appears unresolved.  The He{\sc ii}
emission is well centered on the continuum emission from the ULX, but is
somewhat broader.  We estimate a spatial extent for the He{\sc iii}
region of roughly $1.3\arcsec$ (full width at half maximum)
corresponding to a diameter of 30~pc for a distance of 4.8~Mpc to NGC
5408 \citep{Karachentsev02}.  The H$\beta$ emission is offset from the
continuum and much broader.  We chose to extract spectra from a trace
that is $1.75\arcsec$ wide covering the continuum emission and the
central part of the He{\sc ii} emission.  The trace is 7 pixels wide as
marked by the horizontal line on Fig.~\ref{profile}.  For comparison, we
also examined the spectrum from a $4.5\arcsec$ wide trace.  The H$\beta$
fluxes increases by a factor of 2.0 and the He{\sc ii} $\lambda$4686
increases by a factor of 1.3 for the wide trace.  Since the nebula
appears to have a larger angular size than our slit size ($0.51\arcsec$)
and the background region selected (due to the presence of star A) may
contain some nebular emission, the fluxes quoted here may underestimate
the true fluxes.

Using the Balmer decrements of H$\gamma$/H$\beta$ and H$\delta$/H$\beta$
to estimate the reddening, we find E(B-V) = 0.08$\pm$0.03.  The
reddening estimated from dust within the Milky Way is $\rm E(B-V) =
0.068$ \citep{Schlegel98} and we take this as a lower limit on the
reddening.  The column density measured via the X-ray spectrum is
equivalent to an optical reddening E(B-V) = 0.13 \citep{Predehl95}. If
the optical extinction measures to the edge of a nebula surrounding the
ULX and the additional extinction seen in the X-rays is due to
absorption in the nebula, then the column density through the nebula
would be $N_H = 2.4 \times 10^{20} \rm \, cm^{-2}$.  We corrected for
reddening using the extinction curve from \citet{Cardelli89} with $\rm
R_V = 3.1$.

The combined spectrum from all of the observations is shown in
Fig.~\ref{specall}.  The spectrum is dominated by emission lines, but
continuum emission is also detected.  Table~\ref{lines_table} gives the
dereddened line fluxes relative to H$\beta$ and their widths.  The
dereddened H$\beta$ flux was $(8.6 \pm 0.9)\times 10^{-16} \, \rm erg \,
cm^{-2} \, s^{-1}$.  The widths of the lines are all consistent with the
instrumental resolution except for [O~II] $\lambda\lambda$3726, 3729
which is a blend, He~II $\lambda$4686, and possibly [Ne~V]
$\lambda$3426.

\begin{figure}
\centerline{\includegraphics[angle=0,width=3.0in]{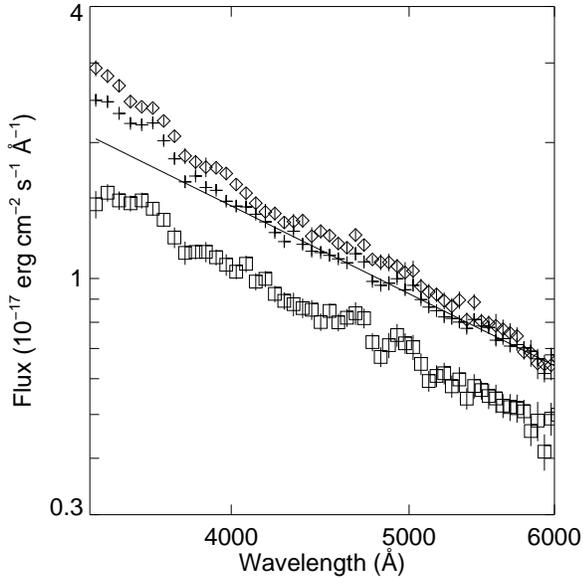}}
\caption{Continuum spectra of the optical counterpart of NGC 5408 X-1. 
The dereddened flux is plotted versus redshift-corrected wavelength. 
The crosses show the average spectrum, the diamonds are for OB6 that had
the highest flux, and the squares are for OB4 that had the lowest flux. 
The solid line is a power-law fitted to the average spectrum in
4000-6000\AA\ range.  The data lie above the fit at wavelengths
shortward of the Balmer edge, indicating a Balmer excess that may be due
to nebular emission.}  \label{contob}  \end{figure}

\begin{deluxetable}{lrrrrrr} \tabletypesize{\scriptsize}
\tablecaption{Emission line and continuum fluxes \label{linevar}}
\tablewidth{0pt} 
\tablehead{\colhead{Line} & \colhead{OB1} & \colhead{OB2} &
                            \colhead{OB3} & \colhead{OB4} &
			    \colhead{OB5} & \colhead{OB6}}
\startdata
$[$O~III] $\lambda$5007 & 223 & 213 & 206 & 219 & 228 & 216 \\
H$\beta$ $\lambda$4861  &  76 &  70 &  70 &  73 &  75 &  72 \\
He~II $\lambda$4686     &  25 &  25 &  24 &  17 &  26 &  25 \\
$[$Ne~V] $\lambda$3426  &  18 &  19 &  18 &  11 &  18 &  20 \\
Continuum               & 1.9 & 2.0 & 1.8 & 1.4 & 2.0 & 2.1 \\
\enddata

\vspace{-2pt}\tablecomments{The table lists the observed fluxes of
four selected emission lines in units of $10^{-17} \, \rm erg \,
cm^{-2} \, s^{-1}$ and the continuum flux integrated over the band
4000-6000~\AA\ in units of $10^{-14} \, \rm erg \, cm^{-2} \, s^{-1}$
for each observation block (OB).}  \end{deluxetable}

We analyzed each observation block individually.  Most of the line
fluxes appear constant with variations on the order of 5\% that may be
due to differences in the placement of the slit.  However, the fluxes
for He~II $\lambda$4686 and [Ne~V] $\lambda$3426 in OB 4 are only 70\%
of the average in the other observations, see Table~\ref{linevar}.  The
average width of the [O~III] $\lambda$5007 line is 5.1\AA.  The average
width of the He~II $\lambda$4686 line, excluding OB4, is somewhat
broader at 6.2\AA.  This suggests an intrinsic width of the He~II line
of 3.5\AA, equivalent to 230~km/s.  The width of the He~II $\lambda$4686
line in OB4 is 5.6\AA, which is marginally narrower than in the in other
observations.

We measured the continuum emission by removing all the detected lines
and then averaging the remaining flux in bins of 50\AA.  We fit a
power-law to the dereddened and redshift-corrected continuum spectrum in
the 4000--6000\AA\ band.  The total flux in this band for each OB is
given in Table~\ref{linevar}.  The level of continuum emission is
roughly constant, within about 5\%, except for OB4 which has about 70\%
of the average flux of the other observations.  The best fitted spectral
index for the average of all observations is $-2.0^{+0.1}_{-0.2}$, where
the uncertainty allows for the uncertainty in the reddening.  The
spectral indexes for the individual OBs are consistent with this range. 
Fig.~\ref{contob} shows the continuum emission for the average spectrum
and the OBs with the highest (OB6) and lowest (OB4) continuum flux.

\section{Discussion}

\subsection{Emission lines}

The fact that the line emission appears spatially extended and the
narrow widths of the lines suggest that they are produced in a nebula. 
However, the variability of the He~II $\lambda$4686 and [Ne~V]
$\lambda$3426 lines suggests that part of the emission in these two
lines is produced close to the compact object, within one light day.

The spectral resolution is not adequate to resolve the [O {\sc ii}]
$\lambda$3729/$\lambda$3726 lines and the coverage does not extend to
the [S {\sc ii}] $\lambda$6716/$\lambda$6731 lines, so these standard
density indicators cannot be used to estimate the density.  From the
flux and spatial extent of the H$\beta$ line, a rough estimate of the
average density is $n_{\rm H} \approx 7 \rm \, cm^{-3}$.  The column
density from the ULX to us through (half of) the nebula is then in
reasonable agreement with the excess absorption in the X-ray spectrum.

The weakness of low-ionization lines, particularly [O {\sc i}]
$\lambda$6300 suggests that shocks are not the dominant source of
ionization.  Using the ratio of [O {\sc iii}] lines, $(f_{\lambda 4959}
+ f_{\lambda 5007})/f_{\lambda 4363} = 53 \pm 7$, we estimate the
nebular temperature to be near 17000~K.  This is consistent with that
expected for photoionization and lower than that expected for shock
ionization \citep{Osterbrock}.  The metallicity indicator ([O {\sc
ii}]($\lambda$3727 + [O {\sc iii}]($\lambda$4959 +
$\lambda$5007))/H$\beta$ = 5.4$\pm$0.2, suggesting a metallicity
somewhat lower than solar \citep{Kewley02}. Using the spectrum from the
$4.5\arcsec$ trace, the same ratio is $5.7 \pm 0.1$.

The strength of the He{\sc ii} $\lambda$4686 and [Ne{\sc v}]
$\lambda$3426 emission is strong evidence that the line emission is
powered by photoionization from a source of high excitation.  The ratio
He{\sc ii}/H$\beta$ equals 0.38 for the average of all observations and
0.27 for OB4.  Both values are higher than found in H{\sc ii} regions in
nearby galaxies \citep{Garnett91} and higher than in the X-ray
photoionized nebula surrounding Holmberg II X-1
\citep{Pakull03,Kaaret04b}.  Within the $1.75\arcsec$ trace in the
average spectrum, the luminosity in He{\sc ii} $\lambda$4686 is $9
\times 10^{35} \rm \, erg \, s^{-1}$ for a distance of 4.8~Mpc to NGC
5408 \citep{Karachentsev02}.  Taking the full extent of the nebula
(within the slit) increases this to $1.2 \times 10^{36} \rm \, erg \,
s^{-1}$.  The [Ne{\sc v}] $\lambda$3426 luminosity integrated over the
full extent of the nebula is $1.0 \times 10^{36} \rm \, erg \, s^{-1}$. 
These values may underestimate the true luminosities, since the nebula
appears to be more spatially extended than our $0.51\arcsec$ slit.  The
variability of the He{\sc ii} line indicates that part of the emission
is due to the compact object instead of the extended nebula.  The He{\sc
ii} luminosity for the wide trace in OB4 is $1.0 \times 10^{36} \rm \,
erg \, s^{-1}$.

The He{\sc ii} emission from NGC 5408 X-1 is a factor of 10 more
luminous than that from the persistent stellar-mass black hole X-ray
binary LMC X-1 \citep{Pakull86} but similar in luminosity and size to
the He{\sc iii} region around the ultraluminous X-ray source Holmberg II
X-1 \citep{Kaaret04b}.  There are powerful radio nebulae associated with
both ULXs \citep{Miller05,Lang07}, while LMC X-1 is radio quiet
\citep{Fender98}.  The higher He{\sc ii} luminosities of the ULXs
suggest that the compact objects produce a greater isotropic luminosity
than does the stellar-mass black hole X-ray binary LMC X-1.  

We used the photoionization code {\it Cloudy} version 07.02
\cite{Ferland98} to model the nebula with the X-ray spectrum described
above as input in order to determine the relation between the He{\sc ii}
luminosity and the ionizing X-ray luminosity.  We placed the X-ray
source at the center of a spherical nebula with  a filling factor of 1
and a radius of 20~pc.  The radius is slightly larger than the observed
size of the He{\sc iii} region since the H$\beta$ emission extends to
lower excitation and thus larger radii.

The relation between He{\sc ii} luminosity and the total luminosity of
the ionizing X-ray/UV radiation is not sensitive to metallicity or
density as long as the density is high enough that the region of fully
ionized He is contained within the simulated nebula and the density is
not so high that there is strong absorption of He{\sc ii} $\lambda$4686
within the nebula.  From the measured He{\sc ii} luminosity for OB4, the
minimum value we observe, we estimated that the true X-ray luminosity
must be at least $2.5 \times 10^{39} \rm \, erg \, s^{-1}$.  This value
is a lower bound since an increased X-ray luminosity would be required
if nebula surrounding the ULX covers only part of the emitted X-rays or
if the radiation field at the outer edge of the nebula is still strong
enough to ionize He.  Given the X-ray flux and spectrum measured with
XMM-Newton, the total intrinsic X-ray luminosity is $1.0 \times 10^{40}
\rm \, erg \, s^{-1}$ and the X-ray luminosity in the observed
0.3-10~keV band is $6.6 \times 10^{39} \rm \, erg \, s^{-1}$ after
correction for absorption.  Thus, the X-ray emission from the ULX is not
strongly beamed.  Use of the [Ne{\sc v}] luminosity produces similar
conclusions, but they are a bit less certain due to the influence of
metallicity.

A {\it Cloudy} simulation with a density of $n_{\rm H} = 6 \rm \,
cm^{-3}$, an input X-ray luminosity of $2.5 \times 10^{39} \rm \, erg \,
s^{-1}$, and a metallicity of $0.5 Z_{\odot}$ produced results that
matched the data for OB4 reasonably well.  In particular, the [O {\sc
iii}] line ratio, $(f_{\lambda 4959} + f_{\lambda 5007})/f_{\lambda
4363} = 54$.  The diameter of the He{\sc iii} region was close to
30~pc.  The He{\sc ii} luminosity was $1.0 \times 10^{36} \rm \, erg \,
s^{-1}$ and the [Ne{\sc v}] luminosity was $8 \times 10^{35} \rm \, erg
\, s^{-1}$.  All of these are close to the observed values for OB4.  The
ratio He{\sc ii}/H$\beta$ = 0.52 is somewhat higher than observed
because our simulated nebula does not cover the full H{\sc ii} region.

We note that the observed H$\beta$ and [O {\sc iii}] emission is offset
from the continuum emission, while the He{\sc ii} emission is reasonably
well centered on the continuum.  Thus, while our {\it Cloudy} simulation
is likely a reasonable approximation to the geometry of the high
excitation regions of the nebula, it does not adequately describe the
more extended, low excitation parts of the nebula.  Therefore, while the
simulation results regarding the He{\sc ii} and [Ne{\sc v}] fluxes  are
likely robust, the results regarding the low excitation lines may be
dependent on the geometry of the nebula.  High resolution imaging of the
nebula in several emission lines will be necessary to understand its
structure in detail.

The nebula is similar in size to the radio nebula near NGC 5408 X-1
\citep{Lang07}.  The ULX Holmberg II X-1 also has both He{\sc ii} and
radio nebulae and their morphology appears to be linked
\citep{Miller05}.  It would be of interest to image both the optical and
radio in more detail and compare their morphologies.

\subsection{Continuum emission}

\begin{figure}
\centerline{\includegraphics[angle=0,width=3.0in]{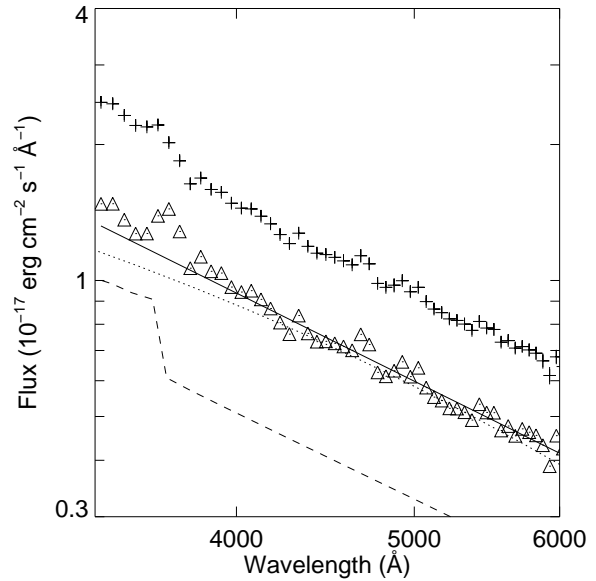}}
\caption{Continuum spectrum of the optical counterpart of NGC 5408 X-1. 
The dereddened flux is plotted versus redshift-corrected wavelength. 
The crosses show the average continuum spectrum.  The dashed line shows
the continuum emission from the photoionized nebula as calculated via
Cloudy.  The triangles show the average continuum spectrum after
subtraction of the nebular component.  The solid line is a power-law
fitted to the nebula-subtracted continuum spectrum in 4000-6000\AA\
range.  The dotted line shows the disk irradiation model fitted to the
X-ray and optical data.}  \label{contcloud}  \end{figure}

Compared with a power-law fitted to the data in the 4000-6000~\AA\ band,
the continuum spectrum, see Fig.~\ref{contob}, shows a marked increase
at short wavelengths with higher flux at wavelengths shorter than the
Balmer edge.  Such a feature is expected from a photoionized nebula. 
The solid line in Fig.~\ref{contcloud} shows the predicted continuum
emission from the Cloudy simulation described below and the triangles
show the data after the nebular continuum is subtracted.  The nebular
continuum matches the enhancement across the Balmer edge reasonably
well.  The excess in the subtracted flux just longward of the Balmer
edge is likely due to high order lines in the Balmer series that the
line removal procedure applied to the optical data does not successfully
identify and remove.

We fitted the continuum after nebular subtraction with a power-law in
the 4000-6000~\AA\ band. The best fitted spectral index is
$-2.01^{+0.07}_{-0.18}$ and the dominant errors are the nebular
subtraction and the reddening correction.  The fitted flux is $8.1
\times 10^{-18} \, \rm erg \, cm^{-2} \, s^{-1} \, $\AA$^{-1}$  at
4300~\AA\ and $4.9 \times 10^{-18} \, \rm photons \, cm^{-2} \, s^{-1}
\, $\AA$^{-1}$  at 5500~\AA.  These translate to Johnson-Cousins
magnitudes of $B_0 = 22.3$ and $V_0 = 22.2$.  The latter value is in
reasonable agreement with that found from HST images of the counterpart,
$V_0 = 22.2$ \citep{Lang07}.  

The X-ray to optical flux ratio, $\xi = B_{0} + 2.5 \log F_{X}$ where
$F_{X}$ is the X-ray flux density in $\mu$Jy after correction for
absorption, is $\xi = 19.6$ at 2~keV  and $\xi = 22.5$ at 0.5~keV. As
noted above, our X-ray and optical observations are separated by 2
years.  However, since the observed X-ray variability of NGC 5408 X-1 is
relatively modest, see \S\ref{xray}, the induced uncertainty in $\xi$ is
only 0.3.  These values of $\xi$ are similar to those found for those
low-mass X-ray binaries where the optical spectra are dominated by
reprocessing of X-rays in the outer accretion disk \citep{vp95}.  The
spectral shape, a power-law with a  spectral index is close to $-2$, is
also similar to X-ray binaries where the optical light is primarily due
to reprocessed X-rays.  Thus, a significant fraction of the light from
the optical companion may arise from reprocessing in the disk.

We fitted the X-ray and optical data simultaneously with the irradiated
disk model of \citet{Gierlinski08}, {\tt diskir}.  We first used the
Cloudy simulation based on the X-ray spectrum modeled with {\tt compps}
to estimate the nebular optical continuum.  We then fitted the X-ray and
optical nebula-subtracted continuum emission data with the {\tt diskir}
model.  We then ran Cloudy again, using this initial set of {\tt diskir}
model parameters.  We used this second Cloudy run to, again, estimate
the nebular optical continuum; this is the dashed line in
Fig.~\ref{contcloud}.  We then performed the nebula subtraction and
fitted the X-ray and optical data with the {\tt diskir} model a second
time.  The parameters from the second iteration of Cloudy are very
similar to those of the first iteration.  The ratio of He{\sc ii}
luminosity to bolometric luminosity of the compact object agrees within
5\% and the other line fluxes agree within 20\%.

Our best fitted irradiated disk model gave a $\chi^2$ of 278.7 for 258
degrees of freedom with a column density $N_H = (6.35 \pm 0.07) \times
10^{20} \rm \, cm^{-2}$, an inner disk temperature $kT = 158 \pm 8 \rm
\, eV$, an asymptotic photon index of $2.36 \pm 0.13$, an electron
temperature of $kT = 1.67^{+0.7}_{-0.3} \pm  \rm \, keV$, and a ratio of
Compton to disk luminosity of $0.66 \pm 0.15$.  The fraction of the
Compton tail thermalized in the disk was fixed at 0.1.  The irradiated
portion of the disk had an outer radius of $8600 \pm 1700$, in units of
the inner disk radius, $r_{\rm in} = 6GM/c^2$, where $M$ is the compact
object mass.  The irradiated disk thermalized a fraction of the 
bolometric luminosity $f_{\rm out} = (6.5 \pm 0.7) \times 10^{-3}$. 
Correcting for absorption and integrating the model from 0.1~eV to
100~keV, the total luminosity is $9.4 \times 10^{39} \rm \, erg \,
s^{-1}$.  As for the Comptonized disk model, the irradiated disk model
required addition of a Gaussian near 0.92~keV to produce an adequate
fit.  We note that the X-ray and optical observations were not
simultaneous.  We redid the fits with the optical fluxes reduced by 30\%
and the only significant change in the parameters was a reduction of
$f_{\rm out}$ to $4.9 \times 10^{-3}$.

The fact that the optical continuum is well described by a model of the
irradiated outer parts of a standard geometrically thin, optically thick
accretion disk raises confidence in the application of standard
accretion disk models to the spectra of NGC 5408 X-1.  The  fraction of
the total luminosity thermalized in the outer disk is higher than seen
from stellar-mass X-ray binaries in the soft or `thermal dominant'
spectral state, but is within in the range seen in states where a
sizable fraction of the emission arises from the Comptonized component
\citep{Gierlinski08}.  The spectral state of NGC 5408 X-1 would be
classified as the `steep power-law state' on the basis of the energy
spectrum and the presence of QPOs in the classification scheme of
\citet{Remillard06}.  It would be of interest to study reprocessing in
stellar-mass X-ray binaries in similar states, in which a moderately
soft power-law component supplies significant flux, to allow a direct
comparison.  From the best fit value found for the outer radius of the
irradiated disk, we calculate an orbital velocity at the outer edge of
the disk of 1200~km/s.  This is larger than the measured width of the
He{\sc ii} emission line, which must arise at least in part from
reprocessing since it is variable.  These results may suggest that we
view the disk close to face-on.

As noted in the introduction, the cool temperatures derived from
standard disk models applied to ULX spectra have been interpreted as
evidence for intermediate mass black holes \citep{Kaaret03}, but the
validity of such models have been ruled out for some sources
\citet{Feng07a}.  The alternative is that the sources are stellar mass
black holes with super-Eddington accretion rates.  In the so-called
`slim disk' models of super-Eddington accretion flows, the disk surface
is convex causing the outer disk to be shielded from the luminous
portions of the inner disk \citep{Abramowicz88}.  Thus, the fraction of
the apparent luminosity that irradiates the outer disk is much lower
than for a standard thin disk.  Super-Eddington accretion may, instead,
cause the ejection of mass leading to a wind that reprocesses the hard
spectrum of radiation from the inner part of the disk into a softer
spectrum \citep{Poutanen07}.  The outer parts of the disk would then see
this softer spectrum, that we measure to have a characteristic
temperature near 0.16~keV.  Such soft X-rays do not thermalize in the
disk and, thus, do not contribution significantly to the reprocessed
emission in the 4000--6000\AA\ band \citep{Gierlinski08}.  Such
accretion flows would thermalize  only a very small fraction of the
apparent luminosity in the outer disk.

The measured ratio of optical to X-ray flux for NGC 5408 X-1 is larger
than would be expected in models of super-Eddington accretion flows, but
is in the range expected from a standard thin disk.  Further
observations of the continuum emission from NGC 5408 X-1 would allow us
to better constrain the properties of the irradiated disk.  Simultaneous
measurement of the optical and X-ray spectrum would provide a better
constraint on the thermalized radiation fraction.  Extension of the
optical coverage into the IR would enable measurement of the roll over
of the irradiated spectrum to a Rayleigh-Jeans spectrum.  From our best
fitted irradiated disk model, this should occur at roughly 1~$\mu$m.
Extension of the optical coverage into the UV could enable measurement
of the long wavelength tail of the direct disk emission.

\section*{Acknowledgments}

This paper is based on observations made with ESO Telescopes at the La
Silla Paranal Observatory under program ID 381.D-0051(A).  We thank the
telescope operators for performing the observations, particularly in
their successful correction of an error in our finding chart.  PK thanks
Sylvain Chaty for his extensive help with the FORS data reduction and
Hua Feng and Ken Gayley for useful discussions.  PK acknowledges partial
support from a Faculty Scholar Award from the University of Iowa and
NASA grant NNX08AJ26G.    SC acknowledges support from the European
Community through Marie Curie FP7 ITN network "Black Hole Universe"
(Grant \#: 215212).  This publication makes use of data products from
the Two Micron All Sky Survey, which is a joint project of the
University of Massachusetts and the Infrared Processing and Analysis
Center/California Institute of Technology, funded by the National
Aeronautics and Space Administration and the National Science
Foundation.



\begin{thebibliography}{}

\bibitem[Abramowicz et al.(1988)]{Abramowicz88} Abramowicz, M.A.,
Czerny, B., Lasota, J.P., Szuszkiewicz, E.\ 1988, ApJ, 332, 646

\bibitem[Appenzeller et al.(1998)]{Appenzeller98} Appenzeller, I.\ et
al.\ 1998, Messenger, 94, 1 

\bibitem[Cardelli, Clayton, and Mathis(1989)]{Cardelli89} Cardelli,
Clayton, and Mathis 1989, ApJ, 345, 245

\bibitem[Dachs, Poetzel, \& Kaiser(1989)]{Dachs89} Dachs, J., Poetzel,
R., Kaiser, D.\ 1989, A\&AS, 78, 487

\bibitem[Fender, Southwell, and Tzioumis(1998)]{Fender98} Fender, R.P.,
Southwell, K., Tzioumis, A.K.\ 1998, MNRAS, 298, 692

\bibitem[Feng \& Kaaret(2005)]{Feng05} Feng, H., \& Kaaret, P.\ 2005,
\apj, 633, 1052

\bibitem[Feng \& Kaaret(2007a)]{Feng07a} Feng, H.\ \& Kaaret, P.\ 2007a,
ApJ, 660, L113 

\bibitem[Feng \& Kaaret(2007b)]{Feng07b} Feng, H.\ \& Kaaret, P.\ 2007b,
ApJ, 668, 941 

\bibitem[Ferland et al.(1998)]{Ferland98} Ferland, G.J., Korista, K.T.,
Verner, D.A., Ferguson, J.W., Kingdon, J.B., Verner, E.M.\ 1998, PASP,
110, 761

\bibitem[Garnett et al.(1991)]{Garnett91} Garnett D.R., Kennicutt R.C.,
Chu Y.-H., Skillman E.D.\ 1991, PASP, 103, 850

\bibitem[Gierli\'nski, Done, \& Page(2008)]{Gierlinski08} Gierli\'nski,
M., Done, C.\ \& Page, K.\ 2008, MNRAS, to appear, arVix:0808.4064

\bibitem[Hutchings et al.(1987)]{Hutchings87} Hutchings, J.B., Crampton,
D., Cowley, A.P., Bianchi, L., Thompson, I.B.\ 1987, AJ, 94, 340

\bibitem[Kaaret et al.(2001)]{Kaaret01} Kaaret, P.\ et al.\ 2001,
MNRAS, 321, L29

\bibitem[Kaaret et al.(2003)]{Kaaret03} Kaaret, P., Corbel, S.,
Prestwich, A.H., Zezas, A.\ 2003, Science,  299, 365.

\bibitem[Kaaret et al.(2004a)]{Kaaret04a} Kaaret, P., Alonso-Herrero, A.,
Gallagher, J.S.\ III, Fabbiano, G., Zezas, A., Rieke, M.J.\ 2004, MNRAS,
348, L28

\bibitem[Kaaret et al.(2004b)]{Kaaret04b} Kaaret, P., Ward, M.\ J., \&
Zezas, A.\ 2004, MNRAS, 351, L83

\bibitem[Karachentsev et al.(2002)]{Karachentsev02} Karachentsev, I.D.\
et al.\ 2002, A\&A, 385, 21

\bibitem[Kewley \& Dopita(2002)]{Kewley02} Kewley, L.J.\ \& Dopita,
M.A.\ 2002, ApJS, 142, 35

\bibitem[Lang et al.(2007)]{Lang07} Lang, C.C., Kaaret, P., Corbel, S.,
Mercer, A.\ 2007, ApJ, 666, 79

\bibitem[Miller et al.(2005)]{Miller05} Miller, N.A, Mushotzky, R.F.,
Neff, S.G.\ 2005, ApJ, 623, L109

\bibitem[Osterbrock \& Ferland(2006)]{Osterbrock} Osterbrock, D.E.\ \&
Ferland, G.J.\ 2006, Astrophysics of Gaseous Nebulae and Active Galactic
Nuclei (2nd ed., University Science Books)

\bibitem[Pakull \& Angebault(1986)]{Pakull86} Pakull M.W., Angebault
L.P.\ 1986, Nature, 322, 511

\bibitem[Pakull \& Mirioni(2003)]{Pakull03} Pakull, M.\ W., \& Mirioni,
L.\ 2002, Revista Mexicana de Astronom\'ia y Astrof\'isica (Serie de
Conferencias) 15, 197

\bibitem[Poutanen \& Svensson(1996)]{Poutanen96} Poutanen J., Svensson
R.\ 1996, ApJ, 470, 249

\bibitem[Poutanen et al.(2007)]{Poutanen07} Poutanen, J., Lipunova, G.,
Fabrika, S., Butkevich, A.G., Abolmasov, P.\ 2007, MNRAS, 377, 1187

\bibitem[Predehl \& Schmitt(1995)]{Predehl95} Predehl, P.\ \& Schmitt,
J.H.M.M.\ 1995, A\&A, 293, 889

\bibitem[Remillard \& McClintock(2006)]{Remillard06}  Remillard, R.~A.,
\& McClintock, J.~E.\ 2006, \araa, 44, 49

\bibitem[Rodr\'{\i}guez, Mirabel, Mart\'{\i}(1992)]{Rodriguez92}
Rodr\'{\i}guez, L.F., Mirabel, I.F., Mart\'{\i}, J.\ 1992, ApJ, 401, L15

\bibitem[Schlegel, Finkbeiner, \& Davis(1998)]{Schlegel98} Schlegel,
D., Finkbeiner, D., \& Davis, M.\ 1998, ApJ, 500, 525

\bibitem[Skrutskie et al.(2006)]{2MASS} Skrutskie, R.M.\ et al.\ 2006,
AJ, 131, 1163.

\bibitem[Strohmayer \& Mushotzky(2003)]{Strohmayer03} Strohmayer, T.E.\
\& Mushotzky, R.F.\ 2003, ApJ, 586, L61

\bibitem[Strohmayer et al.(2007)]{Strohmayer07} Strohmayer, T.\ et al.\
2007, ApJ, 660, 580

\bibitem[Tody(1993)]{Tody93} Tody, D.\ 1993, ``IRAF in the Ninetie'' in
Astronomical Data Analysis Software and Systems II, A.S.P.\ Conference
Ser., Vol 52, eds.\ R.J.\ Hanisch, R.J.V.\ Brissenden, J.\ Barnes, 173.

\bibitem[van Paradijs \& McClintock(1995)]{vp95} van Paradijs, J.\ \&
McClintock J.E.\ in X-Ray Binaries, eds.\ W.H.G.\ Lewin,  J.\ van
Paradijs, \& E.P.J. van den Heuvel, (Cambridge Univ.\ Press, 1995),
pp.\ 58-125.

\end{thebibliography}
\end{document}